\documentclass[11pt,a4paper]{article}
\usepackage{jheppub}
\usepackage{epstopdf}

\usepackage{amssymb}
\usepackage{graphics,bm}
\usepackage{epsfig}
\usepackage{graphicx}
\usepackage{rotating}
\usepackage{color}

\newcommand{\beq}{\begin{equation}}
\newcommand{\eeq}{\end{equation}}
\newcommand{\bqa}{\begin{eqnarray}}
\newcommand{\eqa}{\end{eqnarray}}

\def\square{\vcenter{\vbox{\hrule height.4pt
          \hbox{\vrule width.4pt height4pt
          \kern4pt\vrule width.3pt}\hrule height.4pt}}}

\title{Chiral and deconfinement 
transitions in a magnetic background using the functional renormalization group
with the Polyakov loop}

\author[a]{Jens O. Andersen}
\date{today}
\affiliation[a]{Department of Physics, 
Norwegian University of Science and Technology, 
H{\o}gskoleringen 5,
N-7491 Trondheim, Norway}
\emailAdd{andersen@tf.phys.ntnu.no}
\emailAdd{william.naylor@ntnu.no}
\author[a]{William R. Naylor}
\author[b]{Anders Tranberg}
\emailAdd{anders.tranberg@uis.no}
\affiliation[b]{
Faculty of Science and Technology, University of Stavanger,
N-4036 Stavanger, Norway}
\abstract{
We use the Polyakov loop coupled quark-meson model 
to approximate low energy 
QCD and
present results for the chiral and deconfinement transitions 
in the presence of a constant magnetic background $B$ at finite temperature $T$
and baryon chemical potential $\mu_B$. We investigate effects of various gluonic
potentials on the
deconfinement transition with and without a fermionic backreaction at
finite $B$.
Additionally we investigate the effect of the Polyakov loop on the
chiral phase transition, finding that magnetic catalysis at
low $\mu_B$ is present, but weakened by the Polyakov loop.

}

\begin{document}

\maketitle

\section{Introduction}
\label{sec:introduction}
Knowledge of the equation of state and the phase diagram of QCD is
essential in understanding the properties of the deconfined matter
created in heavy-ion collisions as well as the properties of compact
stars and their quark cores.
In non-central heavy-ion collisions, large time-dependent magnetic
fields are generated during the experiment~\cite{harmen,tonev,bzdak}. The maximum 
strength of these magnetic fields is on the order of $10^{19}$ Gauss ($qB \sim$ 6 $m_\pi^2$).
Likewise, very strong magnetic fields exist inside 
magnetars~\cite{neutron}.
These may be several orders of magnitude larger than 
the magnetic fields in 
ordinary neutron stars.
On the surface, the magnetic field may be as strong as $10^{14}-10^{15}$ Gauss
and it could be as strong as $10^{16}-10^{19}$ Gauss in the interior
of the star. This has spurred the interest in strongly interacting matter
at finite temperature, density and magnetic field, see for example 
Ref.~\cite{overview}
for a recent review.

The phase boundary in $(T,\mu_B,B$) space is therefore of great 
interest;
however due to the infamous sign problem, one cannot use the standard
techniques of lattice calculations at finite $\mu_B$.
At zero $\mu_B$ and finite $B$, there is no sign problem and so
one can calculate the phase diagram in the $T,B$ plane using Monte-Carlo
methods. Recent lattice calculations~\cite{budaleik,gunnar}
suggest that for physical quark masses, the transition temperature
for the chiral transition is a decreasing function of the magnetic field
$B$, while for larger values of the quark masses corresponding to
$m_{\pi}\simeq400$ MeV the temperature is an increasing function of 
$B$~\cite{sanf,negro}. The qualitative behavior of the 
transition temperature for physical quark masses in is disagreement with
model calculations using either the (Polyakov-loop extended)
Nambu-Jona-Lasinio ((P)NJL) 
model or the (Polyakov-loop extended) quark-meson model ((P)QM);
In these models, the critical temperature is an 
increasing function of the
magnetic field, see 
e.g.~\cite{fragapol,sadooghi,pnjlgat,pnjlkas,duarte,skokov,anders,grecoleik}.
Possible resolutions to the disagreement have 
been suggested~\cite{bag,res1,nanko,inverse,huang,ferry,orlov}
and we will discuss these at the end of the paper.

In a previous paper~\cite{anders}, two of us used the 
two-flavor three-color quark-meson model and the functional
renormalization group~\cite{wetterich} 
to map out the phase diagram in the $\mu_B-T$ plane for different values
of the magnetic field (see also Refs.~\cite{ebert22,pinto}). 
In the present paper, we add the Polyakov loop 
to the model to 
include certain aspects of confinement~\cite{fukushima,megias}.
In particular, we investigate a set of possible implementations
of the Polyakov loop
and how they effect both the chiral and deconfinement transitions.
In the context of the functional renormalization
group, this was studied in Ref.~\cite{skokov} at zero baryon chemical
potential.

The paper is organized as follows: In Sec.~\ref{sec:quarkmeson} we
briefly discuss
the functional renormalization group implementation of
the quark-meson model in a constant magnetic background.
In Sec.~\ref{sec:poly} we add the Polyakov
loop variable to the model and review the three gluonic potentials 
we have used in this work. Section~\ref{sec:numerics} 
explans the numerical 
implementation and the effects of the various gluonic potentials. 
In Sec.~\ref{sec:results} we discuss our results for the deconfinement and 
chiral transitions. Finally, in Sec.~\ref{sec:conclusion},  we 
summarise the main results and
comment on the disagreement between lattice and model calculations at finite 
$B$ and $\mu_B=0$.

\section{Quark-meson model and the functional renormalization 
group}
\label{sec:quarkmeson}

The quark meson model is the linear sigma model coupled 
to two massless quark flavors via a Yukawa coupling.
The Euclidean Lagrangian for the model is
\bqa\nonumber
{\cal L}&=&
\bar{\psi}\bigg[
\gamma_{\mu}\partial_{\mu}-\mu\gamma_4
+g(\sigma-i\gamma_5{\boldsymbol \tau}\cdot{\boldsymbol \pi})\bigg]\psi
\label{lagra}
+{1\over2}\bigg[
(\partial_{\mu}\sigma)^2
+(\partial_{\mu}{\boldsymbol \pi})^2
\bigg]
+{1\over2}m^2\bigg[\sigma^2+{\boldsymbol \pi}^2\bigg]
\\ &&
+{\lambda\over4}\bigg[
\sigma^2+{\boldsymbol \pi}^2\bigg]^2
-h\sigma\;,
\eqa
where $\sigma$ is the sigma field, ${\boldsymbol \pi}$ denotes the 
neutral and charged pions.
${\boldsymbol \tau}$ are the Pauli matrices,
$\mu={\rm diag}(\mu_u,\mu_d)$ is the quark chemical potential,
where 
$\mu_u$ and $\mu_d$ are the chemical potential for the $u$ and $d$ quarks,
respectively. 
We set $\mu_u=\mu_d$ so that we are working at zero isospin
chemical potential, $\mu_I=\mbox{$1\over2$}(\mu_u-\mu_d)=0$.
The baryon chemical potential is given by $\mu_B=3\mu$. 
The Euclidean $\gamma$ matrices
are given by $\gamma_j=i\gamma^j_M$, $\gamma_4=\gamma^0_M$,
and $\gamma_5=-\gamma^5_M$, where
the index $M$ denotes Minkowski space.
The fermion field is an isospin doublet,
\bqa
\psi=
\left(\begin{array}{c}
u\\
d\\
\end{array}\right)\;.
\label{d0}
\eqa
If $h=0$, Eq.~(\ref{lagra}) is invariant under $O(4)$.
If $h\neq0$, chiral symmetry is explicitly broken, otherwise
it is spontaneously broken in the vacuum. Either way, the symmetry is
reduced to $O(3)$.
Since $SU(2)_L\times SU(2)_R\sim O(4)$ and $SU(2)_V\sim O(3)$,
the quark-meson model incorporates the global symmetries of two-flavor
QCD, whether or not the $SU(2)_A$-symmetry is broken explicitly 
by finite quark masses.

Chiral symmetry is broken in the vacuum by a nonzero
expectation value $\phi$ for the sigma field and we 
make the replacement
\bqa
\sigma\rightarrow\phi+\tilde{\sigma}\;,
\eqa
where $\tilde{\sigma}$ is a quantum fluctuating field.
The tree-level potential then becomes
\bqa
U_{\Lambda}&=&
{1\over2}m^2_{\Lambda}\phi^2
+{\lambda_{\Lambda}\over4}\phi^4
-h\phi\;.
\label{tree}
\eqa
Note that we have introduced a 
subscript $\Lambda$ on $U$,
$m^2$, and $\lambda$, where $\Lambda$ is the ultraviolet cutoff of the
theory. This  
is a reminder that these are unrenormalized quantities\footnote{
The symmetry breaking term
is equivalent to an external field that does not flow and 
therefore $h=h_{\Lambda}$.}.

We will follow Wetterich's implementation of the renormalization 
group
ideas based on the effective average action 
$\Gamma_k[\varphi]$~\cite{wetterich}.
This action is a functional of a set of background fields that 
are denoted by $\varphi$. $\Gamma_k[\varphi]$ 
satisfies 
an integro-differential flow equation in the variable $k$, to be specified 
below.
The subscript $k$ indicates that all the modes $p$ between the ultraviolet
cutoff $\Lambda$ of the theory and $k$ have been integrated out.
When $k=\Lambda$ no modes have been integrated out and $\Gamma_{\Lambda}$
equals the classical action $S$. On the other hand, when $k=0$, all the
momentum modes have been integrated out and $\Gamma_0$ equals the full
quantum effective action. The flow equation then describes the flow in the 
space of effective actions as a function of $k$.

In order to implement the renormalization 
group ideas, one introduces a regulator function $R_k(p)$.
The function $R_k(p)$
is 
large for $p < k$ and small for $p > k$ whenever $0 < k < \Lambda$, 
and $R_{\Lambda} (p) =\infty$. These properties ensure
that the modes below $k$ are heavy and decouple, and only the modes between $k$ 
and the UV cutoff $\Lambda$ are light and integrated out.
The choice of regulator function
has been discussed in detail in the literature and some choices are
better than others due both to their analytical and stability 
properties, see for example~\cite{litim}.

The flow equation for the effective action cannot be solved exactly
so one must make tractable and yet physically sound approximations.
The first approximation in a derivative expansion is the local-potential 
approximation
and in this case the flow equation for $\Gamma_k$ reduces to a
flow equation for an effective potential $U_k(\phi)$.
In the case of a constant magnetic field, 
the differential equation for $U_k$ first appeared in~\cite{skokov}
and a derivation can be found in the preceeding work~\cite{anders}. It reads
\bqa\nonumber
\partial_k U_k
&=&
{k^4\over12\pi^2}
\left\{
{1\over\omega_{1,k}}\left[1+2n_B(\omega_{1,k})\right]
+{1\over\omega_{k,2}}\left[1+2n_B(\omega_{2,k})\right]
\right\}
\\ &&\nonumber
+k{|qB|\over2\pi^2}\sum_{m=0}^{\infty}{1\over\omega_{1,k}}
\sqrt{k^2-p^2_{\perp}(q,m,0)}\,\theta\left(k^2-p^2_{\perp}(q,m,0)\right)
\left[1+2n_B(\omega_{1,k})\right]
\\ && \nonumber
-{N_c\over2\pi^2}k\sum_{s,f,m=0}^{\infty}
{|q_fB|\over\omega_{q,k}}
\sqrt{k^2-p^2_{\perp}(q_f,m,s)}\,\theta\left(k^2-p^2_{\perp}(q_f,m,s)\right)
\left[1-n^+_F(\omega_{q,k})-n^-_F(\omega_{q,k})
\right]\;,
\\ &&
\label{flowu}
\eqa
where we have defined 
$\omega_{1,k}=\sqrt{k^2+U_k^{\prime}}\,$,
$\omega_{2,k}=\sqrt{k^2+U^{\prime}+2U_k^{\prime\prime}\rho}\,$,
$\omega_{q,k}=\sqrt{k^2+2g^2\rho}\,$,
$p^2_{\perp}(q,m,s)=(2m+1-s)|qB|\,$,
$n_B(x)=1/(e^{\beta x}-1)\,$, $\rho={1\over2}\phi^2$ and 
$n_F^{\pm}(x)=1/(e^{\beta(x\pm\mu)}+1)$, however the fermionic 
distribution functions will be transformed to Eqs.~(\ref{gen1}) and (\ref{gen2})
when we add the Polyakov loop.

At zero temperature, the Bose distribution function vanishes and the 
Fermi distribution function becomes a step function.
Furthermore, if we set $\mu=0$, this step function 
vanishes and we obtain the flow equation in the vacuum.

\section{Adding the Polyakov loop }
\label{sec:poly}
The Polyakov loop $\Phi$ is given by the thermal expectation value of the
trace of the Wilson line, i.e.\ 
\bqa
\Phi={1\over N_c}\langle
{\rm Tr}_c\,L\rangle\;,
\eqa
where the trace is in color space and
\begin{equation}
	L = {\cal P} \exp\left[i\int_0^{\beta}
d\tau \,A_4
\right]\;,
\end{equation}
where $A_4=iA_0$ and $A_0=\delta_{\mu0}{\cal A}_a^{\mu}t^a$.
Here ${\cal A}_a^{\mu}$ are the $SU(3)_c$ gauge fields and
the generators are $t^a={1\over2}\lambda^a$,
where $\lambda^a$ are the Gell-Mann matricies. 
The Wilson line is a complex variable and so $\Phi$
is not equal to 
${\bar \Phi}={1\over N_c}\langle{\rm Tr}_c\,L^\dagger\rangle$ in general.
It is known that
$\Phi=\bar{\Phi}$ at mean field level, but in the present work this is
only true at zero baryon chemical potential.
The Polyakov loop is an order parameter for deconfinement in pure-glue
QCD. Under the center symmetry $Z_N$, it transforms as 
$\Phi\rightarrow e^{2\pi in/N_c}$, where $n=0,1,2...,N_c-1$.
At low temperatures, i.e.\ in the confined phase we have $\Phi\approx0$,
while in the deconfined phase we have $\Phi\approx1$.

Coupling the Polyakov loop to the QM model gives a more physically accurate
model of the quark sector and allows us to explore both the chiral 
and deconfinement transitions of low energy QCD. 
This is done by introducing a constant
background temporal
gauge field ${\cal A}_a^0$ 
via the covariant derivative and adding a phenomenological
potential for the gluonic sector, as discussed below.
The Polyakov gauge is particularly convenient for calculations as 
the Wilson line is then a diagonal matrix, $L=e^{i(\lambda^3A_3+\lambda^8A_8)/2T}$.
Utilizing this and the mean field solution for the effective potential the
quark distribution
functions are found to be transformed from the standard Fermi-Dirac distribution functions to
\begin{align}
\label{gen1}
n_F^+(\Phi, {\bar \Phi}; T, \mu) &= \frac{1+2{\bar \Phi}e^{\beta(E_q-\mu)} 
+ \Phi e^{2\beta(E_q-\mu)}}{1+3{\bar\Phi} e^{\beta(E_q-\mu)}+3\Phi e^{2\beta(E_q-\mu)}
+e^{3\beta(E_q-\mu)}}\;,\\
n_F^-(\Phi, {\bar \Phi}; T, \mu)  &=n_F^+({\bar \Phi}, \Phi; T, -\mu) \; .
\label{gen2}
\end{align}

\noindent These are then substituted back into the
renormalization group flow equation (\ref{flowu}).
This form is a particularly promising result, as in the confining 
limit ($\Phi$ and ${\bar\Phi} \rightarrow 0$) we obtain a Fermi-Dirac-like 
distribution function for states of three quarks, however as $\Phi$ and 
${\bar\Phi} \rightarrow 1$ the functions $n_F^{\pm}$ are equal to the 
standard Fermi-Dirac distribution functions, as they should be.

A number of forms for the gluonic potentials have been proposed 
and investigated at mean field level for the PNJL model~\cite{lourenco} 
and the PQM model with $\mu=0$~\cite{schaefer2010}. In
this work we will investigate three different gluon potentials.
Since the Polyakov loop variable is the order parameter for the $Z(3)$
center symmetry of pure-glue QCD, a Ginzburg-Landau type potential
should incorporate this. A polynomial expansion then leads 
to~\cite{polyakovpot1}
\begin{align}
\frac{U_{\rm poly}}{T^4} &= -\frac{b_2(T)}{2} \Phi{\bar\Phi}  -\frac{b_3}{6}
\big( \Phi^3+{\bar\Phi}^{3} \big) + \frac{b_4}{4} \big( \Phi{\bar\Phi} \big)^2 
\; , \label{poly}
\end{align}
where the coefficients are
\bqa
b_2(T) &=& 6.75 - 1.95\left(\frac{T_0}{T}\right) + 2.624\left(\frac{T_0}
{T}\right)^2 - 7.44\left(\frac{T_0}{T}\right)^3 \; , \\
b_3 &=& 0.75\;,\\
b_4 &=& 7.5\;.
\eqa
The coefficients $b_2(T)$, $b_3$, and $b_4$ are chosen such that the
Polyakov loop potential reproduces the equation of state and 
temperature dependence of $\Phi$ around the transition at $\mu=0$.
The parameter $T_0$ is the transition temperature 
for pure-glue QCD
lattice calculations~\cite{tc270}.

In Refs.~\cite{polyakovpot2,polyakovpot3}, another form for the 
Polyakov loop potential based on the $SU(3)$ Haar measure was proposed:
\begin{align}
	\frac{U_{\rm log}}{T^4} &= -\frac{a(T)}{2} \Phi{\bar\Phi}  +b(T) \ln 
\left[ 1- 6\,{\bar\Phi}\Phi + 4 \big( \Phi^3+{\bar\Phi}^{3} \big) - 
3\big( {\bar\Phi}\Phi \big)^2 \right] \; , \label{log}
\end{align}
where the coefficients are
\begin{align}
	a(T) &= 3.51 - 2.47\left(\frac{T_0}{T}\right) + 15.2\left(\frac{T_0}{T}
\right)^2 \; , \\
	b(T) &= -1.75\left(\frac{T_0}{T}\right)^3\;.
\end{align}
We note that the logarithmic term ensures that the magnitude of 
$\Phi$ and $\bar{\Phi}$ are constrained to be in the region between 
$-1$ and $1$, i.e.\ the possible attainable 
values for the
normalized trace of an element of the $SU(3)$.
Finally, Fukushima proposed a Polyakov loop potential in~\cite{polyakovpot4}
\begin{align}
\frac{U_{\rm Fuku}}{T^4} &= -\frac{b}{T^3} \left( 54e^{-a\,T_0/T}\Phi{\bar\Phi}
  +\ln \left[ 1- 6\,\Phi{\bar\Phi} + 4 \big( \Phi^3+{\bar\Phi}^{3} \big) 
- 3\big( \Phi{\bar\Phi} \big)^2 \right] \right)\; , \label{Fuku}
\end{align}
where the constants are $a=664/270$ and $b=(196.2\textrm{ MeV})^3$ and we
have added dependence upon the transition temperature, $T_0$.

A problem with all the Polyakov loop potentials proposed is that
they are independent of the number of flavors and of the baryon chemical
potential. However, we know that, for example, the transition temperature
for the deconfinement transition is a function of $N_f$.
In other words, one ought to incorporate the back-reaction from the 
fermions to the gluonic sector. In Ref.~\cite{bj}, the authors
use perturbative arguments to estimate the effects of the number
of flavors and the baryon chemical potential on the transition temperature
$T_0$. The functional form of $T_0$ is~\cite{herbst}
\begin{equation}
	T_0 = T_\tau e^{-1/(\alpha_0 \, b(N_f,\mu))}\;,
\label{nfmub}
\end{equation}
where
\begin{equation}
b(N_f,\mu) = \frac{1}{6\pi}(11N_c - 2N_f) - \frac{16}{\pi}N_f \frac{\mu^2}
{(\hat{\gamma} \, T_\tau)^2}\;,
\label{bpar}
\end{equation}
and $T_\tau = 1.77 \textrm{ GeV}$, $\alpha_0 = 0.304$. ${\hat\gamma}$ controls 
the curvature of $T_0$ as a function of $\mu$, and again 
following~\cite{herbst} we experiment with a range of values to study the
effects. This is further discussed in the following section.

Let us finally make a few remarks about the sign problem.
At finite baryon chemical potential, QCD has a sign problem due to 
a complex fermion determinant. This implies that the action is complex
and one cannot use standard Monte-Carlo techniques based on importance 
sampling. Also effective models 
such as the PNJL and PQM models have a sign problem at finite
baryon chemical potential as discussed in 
Refs.~\cite{sign1,sign2,sign3,pbarp} for example. The sign problem in these 
models
shows up as an imaginary part of the effective potential and one must
therefore consider it as a complex function of complex 
variables $\Phi$ and $\bar{\Phi}$. 
There are two ways out. One way is to restrict the
Polyakov loop variables to be real as in Ref.~\cite{fukushima}.
This is the approach we will follow in the present paper.
The other option is to split the effective potential into a real part
and an imaginary part~\cite{sign3,pbarp} and treat the imaginary part
as a perturbation. While this is no longer the case when including
perturbative corrections, at the mean-field level this implies $\Phi=\bar{\Phi}$.

\section{Numerical implementation and the glue potential}
\label{sec:numerics}

To find the equilibrium state values of the order parameters
$\phi$, $\Phi$ and ${\bar\Phi}$ we numerically solve the flow equation
(\ref{flowu}) with the boundary condition specified by the tree level
potential, Eq.~(\ref{tree}), on a grid in  $\phi$-$\Phi$-${\bar\Phi}$-space
with $\phi \in [0,126]$ MeV and $\Phi , {\bar\Phi} \in [0,1]$ ($\Phi$
and ${\bar\Phi}$ are real, as discussed Sec.~\ref{sec:poly}).
Doing this at various values of $T$, $B$ and $\mu$ gives
us $U_{k=0}(\phi,\Phi,{\bar\Phi};T,B,\mu)$, which we construct as a dimensionless quantity.
In the derivation of the flow
equation we have used $O(4)$ symmetry, thus for the boundary condition of
the flow we set $h=0$, then when minimising with respect to $\phi$ we
minimise $U_{k=0} - h\phi$. The resulting surface, $U_{k=0}(\Phi , {\bar\Phi})$
is very smooth thus we use interpolation to save computation time. Additional
runs at intermediate values show that errors due to the interpolation are
on the order of 0.1\%. Before we minimise with respect to the deconfinement
order parameters we must add the gluonic potential. Thus $\Phi$
and ${\bar\Phi}$ are obtained from the minimisation of
$U_{k=0}(\Phi , {\bar\Phi}) + U_{\textrm{glue}}(\Phi , {\bar\Phi})/\Lambda^4$,
where `glue' stands for one of `poly', `log' or `Fuku' as given in 
Sec.~\ref{sec:poly}.

We use the following (dimensionless) bare parameters: $m_\Lambda^2 = 0.075$, 
$\lambda_\Lambda = 9.2$, $g=3.2258$ and $h=0.0146$ and we have 
$\Lambda=500$~MeV  which 
give constituent quark masses of 300~MeV, a sigma mass of $\sim$478~MeV and pion
masses of $\sim$140~MeV, that is, our results are calculated at the physical 
point. Changing the energy of the ultraviolet cutoff from 500 to 800 MeV, gives an 
increase of approximately 3\% to the chiral phase transition at low $\mu$, and 
approximately 10\% at low $T$.
Additional details about the implementation at $\Phi={\bar\Phi}=1$ can 
be found in~\cite{anders}.

\begin{figure}[htb]
\begin{center}
\setlength{\unitlength}{1mm}
\includegraphics[width=16.0cm]{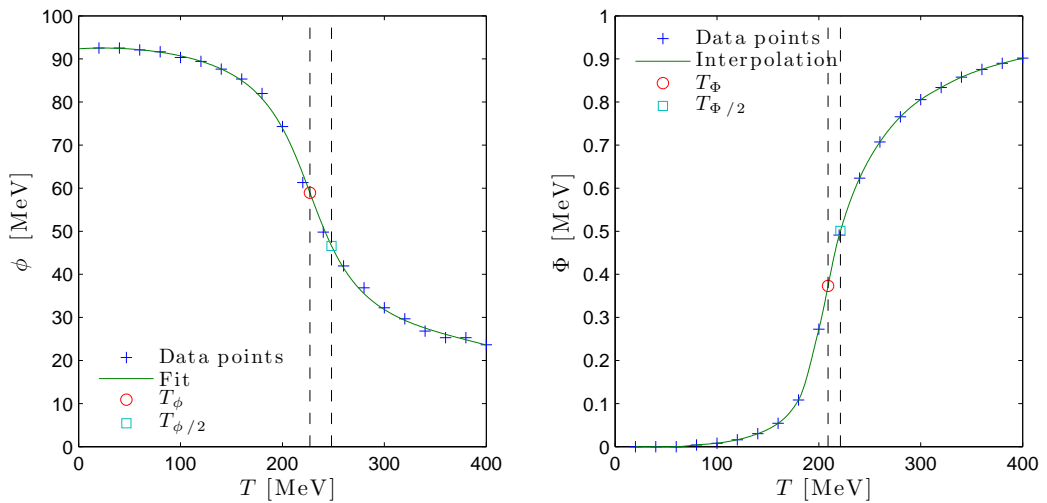}

\caption{Methods used to determine the transition temperatures for
the chiral transition (left) and deconfinement transition (right).
Both plots are for $\mu=0$. See main text for details.}
\label{tcdef}
\end{center}
\end{figure}

\begin{figure}[htb]
\begin{center}
\setlength{\unitlength}{1mm}
\includegraphics[width=16.0cm]{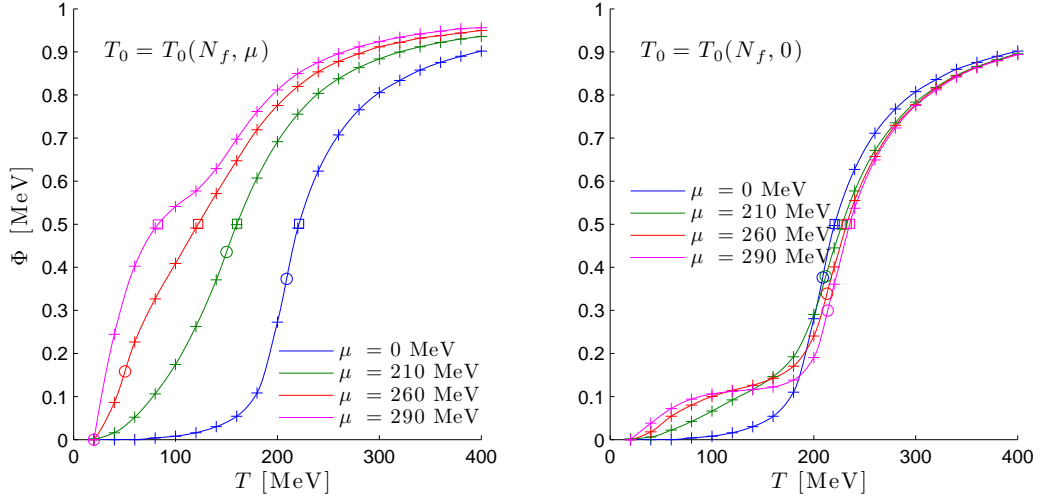}
\caption{Order parameter $\Phi$ as a function of
$T$ for various values of the chemical potential $\mu$ with (left) and
without (right) a $\mu$ dependent gluonic transition temperature, $T_0$.
$+\,$s are the data points, lines are the interpolations thereof,
$\square\,$s give $T_{\Phi/2}$ and $\circ\,$s approximate $T_{\Phi}$}
\label{dcop}
\end{center}
\end{figure}

As the results presented here are calculated at the physical point all of the 
phase transitions are
crossover `transitions' and thus all critical temperatures are
pseudo-critical temperatures.
We must therefore define how we can calculate these transitions.
Since we have discretized the variables in the computation
of the effective potential, calculating the inflection point directly from
the output data is very inaccurate. Thus
one way to define the transition temperature is to fit the data points
for the order parameter in question with a function and then define the 
transition
temperature, $T_{\textrm{x}}$, as the inflection point of the fitted curve.
For the chiral transition we use this method,
with the fit based on $\arctan(x)$.
However, using this method for the deconfinement transition we run into
problems as the functional form of the underlying curve changes with
changing $\mu$ (see the left panel of Fig.~\ref{dcop}). An alternative way of 
defining this transition is
when the order parameter, $\Phi(T)$, is equal to ${1\over2}$, this we define 
as $T_{\Phi/2}$.
To find this point we interpolate with third-order polynomial interpolation.
Fig~\ref{tcdef} illustrates this for $\mu=0$.
The left panel shows the data points (crosses) for $\phi$
as a function of $T$. The  open circle indicates the inflection point
of the fitted curve, i.e. $T_{\phi}$, while the cross indicates the
temperature when the normalized chiral order parameter satisfies
$\phi/\phi(T=0)={1\over2}$, we denote this $T_{\phi/2}$.
The right panel shows the same, but now for the deconfinement 
order parameter $\Phi$ and the green curve is now the interpolation used to 
determine $T_{\Phi/2}$.

\begin{figure}[htb]
\begin{center}
\setlength{\unitlength}{1mm}
\includegraphics[width=12.0cm]{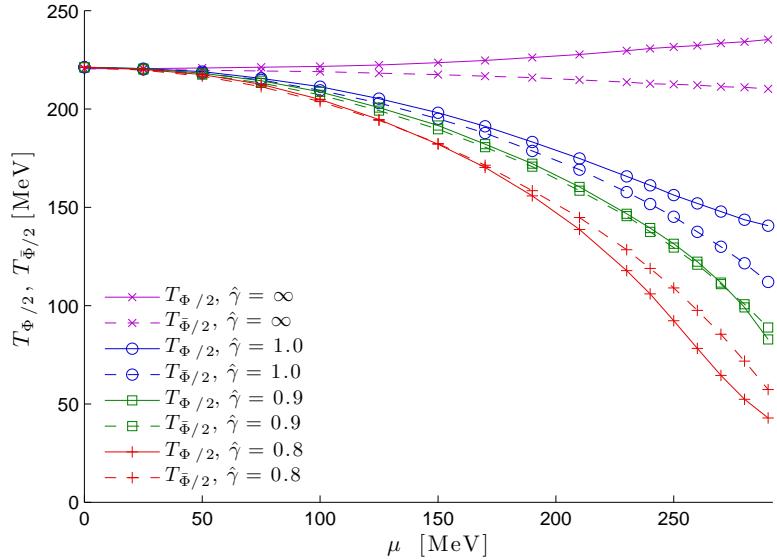}
\caption{Phase diagram for the deconfinement transition with 
$B=0$ and various values of the parameter
$\hat{\gamma}$. See main text for details.}
\label{dcgamma}
\end{center}
\end{figure}

Following Ref.~\cite{bj}, we introduced an $N_f$ and $\mu_B$-dependent
transition temperature $T_0$ via Eq.~(\ref{nfmub}).
In Fig.~\ref{dcgamma}, we show the effects of varying the parameter
$\hat{\gamma}$ in Eq.~(\ref{bpar}) on the deconfinement transition
in the $\mu-T$ plane for zero magnetic field and utilizing the polynomial gluonic potential, Eq.~(\ref{poly}). The solid lines show
$T_{\Phi/2}$
while the dashed lines show $T_{{\bar\Phi}/2}$
for the same values of $\hat{\gamma}$. We note that both $\Phi$
and $\bar{\Phi}$ are real and coincide for $\mu=0$
but differ at non-zero $\mu$.
Furthermore, for a $\mu_B$-independent $T_0$
($=208$ MeV) 
the transition temperature
is almost independent of the baryon chemical potential $\mu$
(magenta lines).
The red, green, and blue lines show the results for $\hat{\gamma}=0.8$,
$0.9$, and $1.0$, respectively. The bending of the curves decreases
as a function of $\hat{\gamma}$ which is reasonable since this parameter enters
in the denominator the parametrization (\ref{bpar}) of $b(N_f,\mu)$.
Finally, we remark that the qualitative behavior is the same for finite
magnetic field $B$. We will present more results for various $B$-fields
in the next section.

In Fig.~\ref{dcop}, we show
the order parameter $\Phi(T)$ as a function of
$T$ for $\mu=0$ (blue), $\mu=210$ (green), 
$\mu=260$ (red), and $\mu=290$ (magenta) with and without
a $\mu$ dependent gluonic potential.
In the left panel, the results are for $T_0=T_0(N_f,\mu)$,
while in the right panel $T_0=T_0(N_f,0)$ i.e.\ independence
from $\mu$.
Comparing the two panels we see the result shown in Fig.~\ref{dcgamma},
that only with a $\mu$ dependent transition temperature $T_0$ do we obtain
significant change in the deconfinement order parameter when varying $\mu$.
Additionally we see in the right panel that at high $\mu$ (magenta in 
particular)
there is an initial increase in $\Phi$ around $T=50$ MeV, which comes from the 
mesonic and fermionic
potential, $U_{k=0}$, and then around 208 MeV there is the typical increase, 
driven
largely by the gluonic potential, $U_{\textrm{glue}}$. We then see in the left 
panel, with
a $\mu$ dependent $T_0$, that the effect of $U_{\textrm{glue}}$ mirrors that of 
$U_{k=0}$
and the deconfinement transition thus decreases with increasing $\mu$.

Figure~\ref{dcop} also illustrates the aforementioned difficulties 
in defining the deconfinement transition at large $\mu$. It is seen
that $T_{\Phi/2} \sim T_\Phi$
at low $\mu$, but for $\mu \gtrsim 230$ MeV this is no longer true.
In addition to this,
the numerics become more time consuming at low $T$,
thus for values of $T \gtrsim 30$ MeV our results
only approximate the behavior of the model.
For these reasons we have only calculated the
phase diagram up to $\mu=290$ MeV.
\footnote{We
have also observed the splitting of the chiral transition reported
in~\cite{herbst13} without the Polyakov loop, but have not
resolved that region in detail with the Polyakov loop.}


\begin{figure}[htb]
\begin{center}
\setlength{\unitlength}{1mm}
\includegraphics[width=12.0cm]{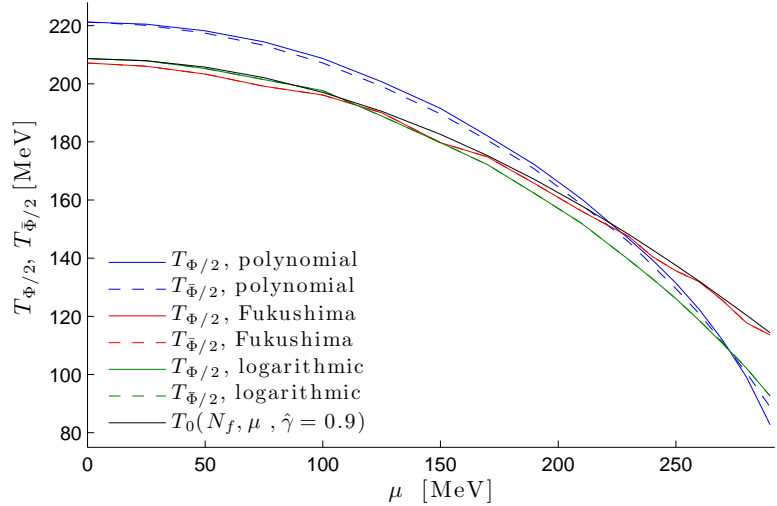}
\caption{Phase diagram for the deconfinement transition for different 
glue potentials and $B=0$. 
Also shown is the transition temperature 
$T_0=T_0(N_f,\mu,\hat{\gamma}=0.9)$ for pure glue for comparison.}
\label{gluepot}
\end{center}
\end{figure}

In Fig.~\ref{gluepot}, we show the phase diagram for the deconfinement
transition with the three different glue potentials introduced in Sec.~\ref{sec:poly} at $B=0$. The blue lines are the
polynomial potential~(\ref{poly}), the red lines are the 
logarithmic potential~(\ref{log}), and the green lines are the 
Fukushima potential~(\ref{Fuku}). 
The black line shows 
the transition temperature $T_0=T_0(N_f,\mu,\hat{\gamma}=0.9)$
for pure glue for comparison.
We note that the black curve is almost the same as the
curve for the Fukushima potential (red), implying that the coupling to 
the quarks has almost no influence on the deconfinement transition.

As was observed in~\cite{polyakovpot2} we find with the
logarithmic potential that $\Phi=\bar{\Phi}$ for all values of $\mu$,
we also find this to be true with the Fukushima potential. We also find
with the Fukushima potential, and to a lesser degree with the logarithmic 
potential,
that the deconfinement transition temperature is dominated by the gluonic 
potential.
This was also backed up by direct investigation of the $\Phi$ and ${\bar\Phi}$ 
as functions of $T$.


\begin{figure}[htb]
\begin{center}
\setlength{\unitlength}{1mm}
\includegraphics[width=12.0cm]{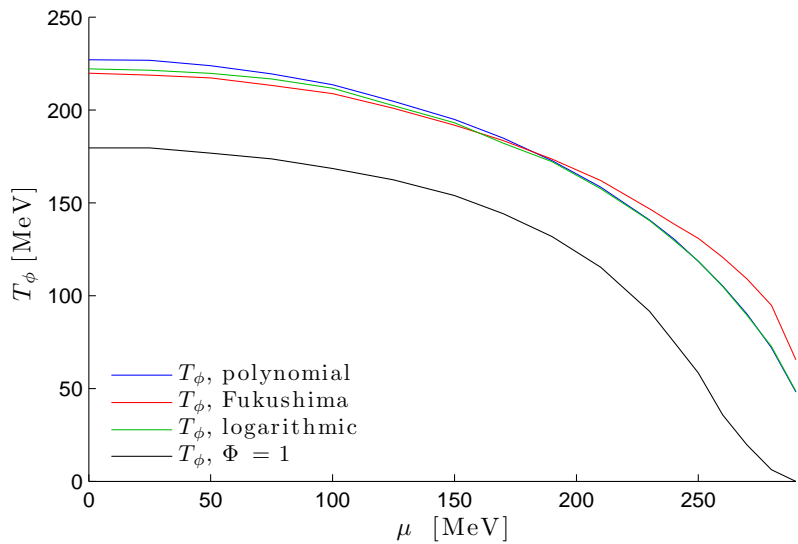}
\caption{Phase diagram for the chiral transition for different 
glue potentials and $B=0$. 
Also shown is the transition temperature for $\Phi=1$, i.e.\ for
the quark-meson model without the Polyakov loop.}
\label{gluepot2}
\end{center}
\end{figure}

In Fig.~\ref{gluepot2}, we show the phase diagram for the chiral transition
using the different gluonic potentials. We also show the phase diagram
for the quark-meson model without the Polyakov loop, i.e. for
$\Phi=1$.
The lines show that the particular form of the gluonic potential 
is 
not as influential as we saw in the case of the deconfinement transition.
At zero $\mu$ and $B$, $T_\phi$ decreases by 2\% and 3\%
for the logarithmic and Fukushima potentials respectively. Only with
$\mu \gtrsim 260$ MeV do we see a significantly larger deviation than this.

\newpage

\section{Results at finite magnetic field}
\label{sec:results}
\begin{figure}[htb]
\begin{center}
\setlength{\unitlength}{1mm}
\includegraphics[width=12.0cm]{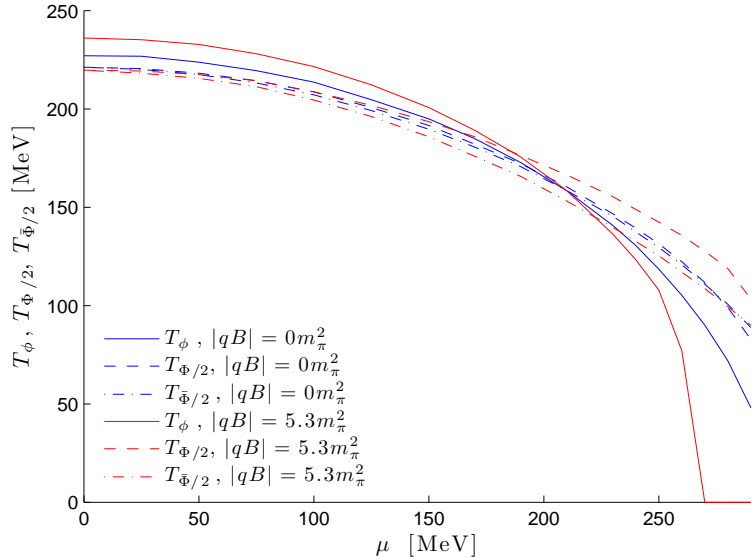}
\caption{Phase diagram for 
the deconfinement and chiral transitions for
$B=0$ and the largest magnetic
field, $|qB|=5.3$ $m_{\pi}^2$ with the Polynomial potential.}
\label{phase}
\end{center}
\end{figure}

\begin{figure}[htb]
\begin{center}
\setlength{\unitlength}{1mm}
\includegraphics[width=15.0cm]{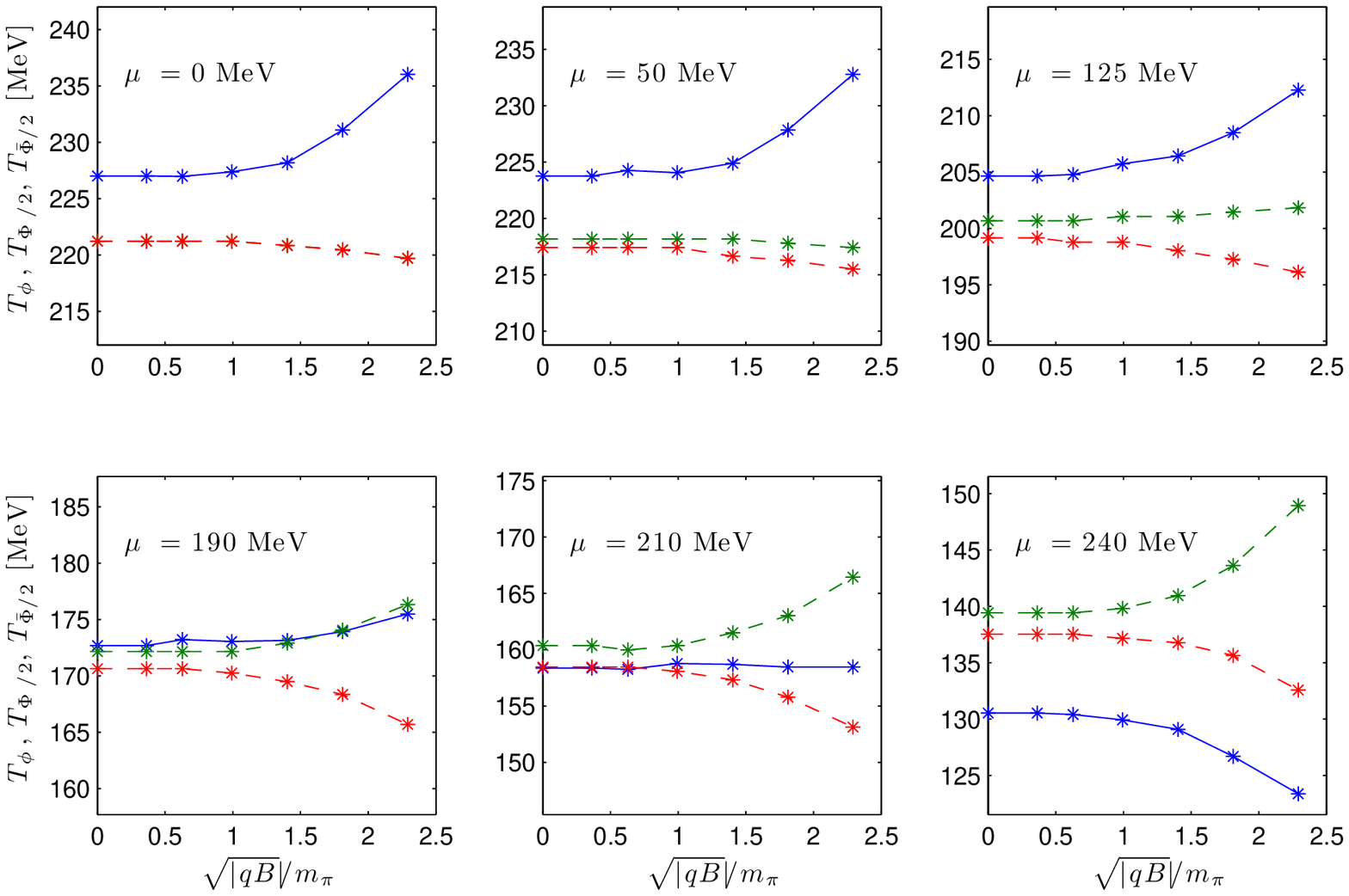}
\caption{Transition temperatures for the chiral and deconfinement transitions
as functions of $B$ for different values of $\mu$. Solid blue lines denote $T_\phi$ while dashed lines correspond to the deconfinement transition with green giving $T_{\Phi/2}$, red giving $T_{{\bar\Phi}/2}$.}
\label{big}
\end{center}
\end{figure}

In this section, we will present our main results and discuss them in some
detail. In Fig.~\ref{phase}, we show the phase diagram for the
chiral and the deconfinement transitions for $B=0$ (blue lines)
and for $|qB|=5.3m_{\pi}^2$. The results are obtained using the
polynomial glue potential~(\ref{poly}). We will discuss the
results in detail in connection with Fig.~\ref{big}, where we
show the chiral and deconfinement transition temperatures as a function of $B$
for different values of $\mu$.

In Fig.~\ref{big}, we show the 
transition temperatures for the chiral and deconfinement transitions
as functions of $B$ for different values of $\mu$.
The solid blue lines indicate the chiral transition, $T_\phi$,
while the dashed green lines are $T_{\Phi/2}$
and the dashed red lines are $T_{{\bar\Phi}/2}$.
In the left upper panel, $\mu=0$ and $\Phi={1\over2}$ and $\bar{\Phi}={1\over2}$ coincide
for all B,
in agreement with our earlier remarks about the sign problem.
We note that the transition temperature for the chiral transition 
is increasing for values of $\mu$ up to approximately $\mu=210$ MeV where
it is flat (lower middle panel). For larger chemical potentials,
the transition temperature for chiral transition is a decreasing function.
This shows the magnetic catalysis for small $\mu$ and inverse catalysis
for large $\mu$ which we discuss below.
For nonzero $\mu$ we see that the splitting between $\Phi$ and $\bar{\Phi}$
increases with $\mu$ and also with the strength of the magnetic field $B$.
For small values of $\mu$, $T_{\Phi/2}$ and
$T_{{\bar\Phi}/2}$ are almost independent of
$B$, while for large values, $T_{\Phi/2}$ increases with
increasing $B$ while $T_{{\bar\Phi}/2}$ decreases with $B$.
This behavior indicates that the relative importance
of the fermionic and mesonic fields also increases with larger $B$ and $\mu$ 
although we have not identified a mechanism behind this behavior.

\begin{figure}[htb]
\begin{center}
\setlength{\unitlength}{1mm}
\includegraphics[width=12.0cm]{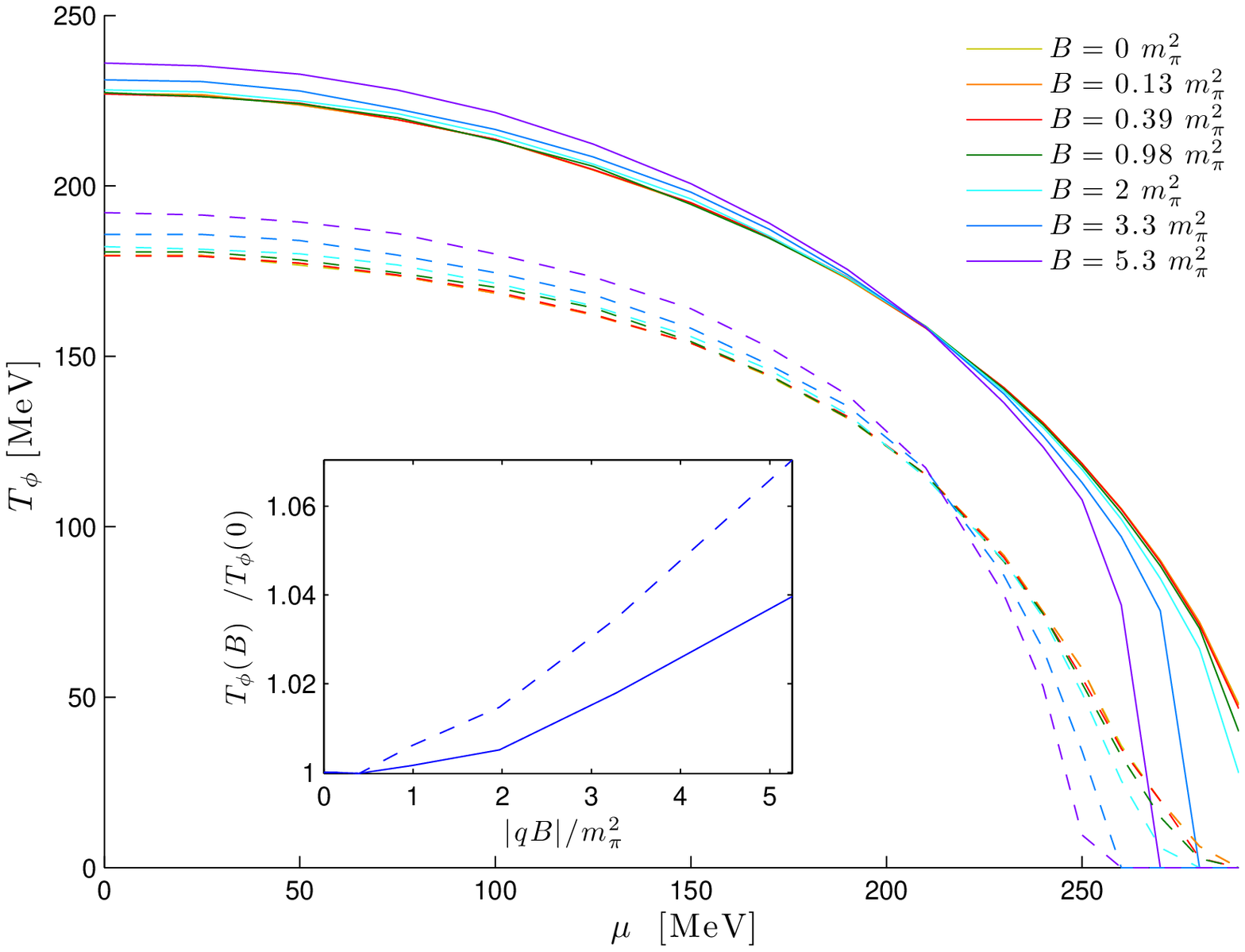}
\caption{Phase diagram for the chiral transition for different values of
the magnetic field $B$ with (solid lines) and without (dashed lines)
the Polyakov loop using the polynomial potential. 
Inset shows the critical temperature as a function of $B$
for $\mu=0$ with (solid lines) and without (dashed lines) the Polyakov loop.}
\label{cata}
\end{center}
\end{figure}

In Fig.~\ref{cata}, we show the phase diagram for the chiral phase
transition for different values of the magnetic field $B$ with 
coupling to the Polyakov loop variable (solid lines) and without 
(dashed lines). Inset shows the transition temperature as a function
of $B$ for vanishing $\mu$ in the two cases.
We first notice that the critical temperature 
increases with the magnetic field for small values of the chemical
potential $\mu$.
The basic mechanism is that of magnetic catalysis~\cite{dcsm1,dcsm2,catarev},
namely that the chiral condensate increases as a function of 
the magnetic field. It is interesting to note that the
increase of the transition temperature 
as a function of $B$ is smaller when we couple the chiral sector
to the gluonic sector.
For large values of the chemical potential
$\mu$, the critical temperature is a decreasing function of
the magnetic field. This is inverse catalysis~\cite{inverse0,inverse2}.
We also find that the transition temperature is increased signficantly for
all values of $\mu$ with the addition of the Polyakov loop. Below $\mu\sim200$ 
MeV $T_\phi$ increases by approximately
25\% and above this density we find greater
increases in $T_\phi$. The Polyakov loop acts to suppress the finite temperature,
fermionic
contribution to the effective potential at all temperatures, although 
particularly at
low temperatures. Thus we expect some increase in $T_\phi$ but its magnitude is 
of
interest as it shows that the confining dynamics does play an important role in
the chiral transition within this model. 
In this region we find $T_{\phi\textrm{, Fuku}} - T_{\phi\textrm{, log/poly}} \approx 20$ 
MeV.
The relative increase in magnetic field is more greatly affected by the choice 
of potential,
with the relative increase in $T_\phi$ being approximately 20\% less with the 
logarithmic and 
Fukushima potentials as opposed to the polynomial potential shown in 
Fig.~\ref{cata}.



Very recently, the existence of a new critical point associated with the
deconfinement transition of strongly interacting matter at finite $T$
and $B$, but vanishing $\mu$ has been suggested~\cite{cohen}.
The basic idea is that quarks effectively decouple in 
the presence of very large magnetic
fields due to their increasing mass as a function of $B$.
In this case, one should be able to describe the system
with an effective theory of pure gluondynamics.
Although this effective theory is anisotropic, it is likely that it
has a first-order transition just like isotropic pure-glue QCD.
Since QCD with physical quark masses exhibit a crossover and not
a first-order transition, there ought to be a critical point
in the $T-B$ plane,
where the line of first-order transition ends.
However we find no evidence
within the range of magnetic fields we examine of a transformation from
the observed cross-over transition to a first order transition for the 
deconfinement order parameter.

\section{Summary and outlook}
\label{sec:conclusion}

In this work we have used the functional renormalization group to calculate
the phase diagram with respect to the chiral and deconfinement transitions
for the Polyakov loop extended quark-meson model. We first investigated the 
effects of the gluonic potential, showing that the
deconfinement transision is quantitavely dependent upon the exact 
implementation,
and in some cases even qualitatively dependent. Most noticibly 
$T_{\Phi/2} - T_{{\bar\Phi}/2}$
is only non-zero when using the polynomial potential~(\ref{poly}). This 
potential was also
the least dominating in that the fermionic and mesonic degrees of freedom had a 
much
larger effect upon the deconfinement order parameters, $\Phi$ and ${\bar\Phi}$. 
However for
all three potentials the gluonic potential dominated the dynamics. At
high $\mu$ we see a double humped structure in these order parameters.
This made the evaluation of $T_{\Phi/2}$ and $T_{{\bar\Phi}/2}$ difficult and
we can not find a first order transition around $\mu\sim300$ MeV (given by
Herbst et al.~\cite{herbst}) although we saw indications of this.

We find magnetic catalysis at
low $\mu$ in agreement with other model calculations,
however we see a weakening
of its effects with the addition of the Polyakov loop. At large $\mu$ the
inverse magnetic catalysis found in the quark-meson model~\cite{anders}
is also found here.
When using the polynomial potential we a find splitting of
$T_{\Phi/2}$ and $T_{{\bar\Phi}/2}$ at non-zero $\mu$. This splitting increases with 
increasing
magnetic field strength and quark chemical potential (other than for the very 
highest $\mu$ value).
In addition
$T_\phi$ increases significantly 
for all values of $\mu$ shows that the Polyakov loop plays an important role
in the chiral transition. In contrast to the confinement transition,
we found that the chiral transition is not sensitive to the choice of
the gluon potential.

In the recent papers~\cite{budaleik,inverse}, the authors 
suggest a resolution
of the discrepancy between the
model calculations and the lattice simulations.
The chiral condensate can be written as
\bqa
\langle\bar{\psi}\psi\rangle
&=&{1\over{\cal Z}(B)}\int d{\cal U}
e^{-S_g}\det(D\!\!\!\!/(B)+m){\rm Tr}
(D\!\!\!\!/(B)+m)^{-1}\;,
\label{det}
\eqa
where the partition function is
\bqa
{\cal Z}(B)&=&\int d{\cal U}
e^{-S_g}\det(D\!\!\!\!/(B)+m)\;,
\eqa
and $S_g$ is the pure-glue action. Thus there are two contributions to the
chiral condensate, namely the operator itself 
(called valence contribution)
and the change of typical gauge configurations sampled, coming
from the determinant in Eq.~(\ref{det})
(called sea contribution). At least for
small magnetic fields one can disentangle these contributions by 
defining
\bqa
\langle\bar{\psi}\psi\rangle^{\rm val}
&=&{1\over{\cal Z}(0)}\int d{\cal U}
e^{-S_g}\det(D\!\!\!\!/(0)+m){\rm Tr}
(D\!\!\!\!/(B)+m)^{-1}\;,
\\
\langle\bar{\psi}\psi\rangle^{\rm sea}
&=&{1\over{\cal Z}(B)}\int d{\cal U}
e^{-S_g}\det(D\!\!\!\!/(B)+m){\rm Tr}
(D\!\!\!\!/(0)+m)^{-1}\;.
\eqa
At zero temperature, both contributions are positive leading
to magnetic catalysis. At temperatures around the transition temperature,
the valence condensate is still positive while the sea condensate
is negative. Hence there is a competition between the two leading to a
net inverse catalysis. The sea contribution can be viewed as a back reaction
of the fermions on the gauge fields and this effect is not present in the 
model calculations as there are no dynamical gauge fields.
If such a back reaction can be incorporated in the model calculations,
one may be able to obtain agreement with the lattice simulations.
One way of doing this is by using a $B$-dependent 
parametrization 
of the
transition temperature~\cite{ferry} 
in analogy with the flavor and $\mu_B$ dependence that we see
here to be critical to realistic mapping of the phase diagram.
We will report on this in a future publication~\cite{future}.

\section*{Acknowledgments}
The authors would like to thank Jonas R. Glesaaen for valuable 
discussions.


\appendix


\end{document}